%% file: conference_101719.tex
\documentclass[conference]{IEEEtran}
\IEEEoverridecommandlockouts
\usepackage{cite}
\usepackage{amsmath,amssymb,amsfonts}
\usepackage{algorithmic}
\usepackage{graphicx}
\usepackage{textcomp}
\usepackage{xcolor}
\usepackage{glossaries}
\usepackage[hyphens]{url}
\usepackage{hyperref}
\usepackage[noabbrev,capitalise]{cleveref}
\usepackage{orcidlink}
\usepackage[separate-uncertainty=true]{siunitx}
\usepackage{booktabs}
\usepackage{multirow}
\usepackage{makecell}
\usepackage{threeparttable}
\usepackage[export]{adjustbox}
\usepackage[caption=false,font=footnotesize]{subfig}
\usepackage{tikz,pgfplots,pgfplotstable}
\usetikzlibrary{patterns, spy, backgrounds}
\pgfplotsset{compat=1.18}
\usepackage{enumitem}
\usepackage[subtle,tracking=normal]{savetrees}

\renewcommand{\baselinestretch}{0.925}

\makeatletter
\newcommand{\linebreakand}{%
  \end{@IEEEauthorhalign}
  \hfill\mbox{}\par
  \mbox{}\hfill\begin{@IEEEauthorhalign}
}
\makeatother

\newcolumntype{H}{>{\setbox0=\hbox\bgroup}c<{\egroup}@{}}

\glsenableentrycount

\newacronym[longplural=systems-on-chip]{soc}{SoC}{system-on-chip}
\newacronym{isa}{ISA}{instruction set architecture}
\newacronym{pulp}{PULP}{parallel ultra-low power}
\newacronym{odrg}{ODRG}{on-demand redundancy grouping}
\newacronym{tcls}{TCLS}{triple-core lockstep}
\newacronym{ecc}{ECC}{error correction codes}
\newacronym{dut}{DUT}{device under test}
\newacronym{wdt}{WDT}{watchdog timer}
\newacronym{seu}{SEU}{single-event upset}
\newacronym{see}{SEE}{single-event effect}
\newacronym{sefi}{SEFI}{single-event functional interrupt}
\newacronym{set}{SET}{single-event transient}
\newacronym{sel}{SEL}{single-event latchup}
\newacronym{mbu}{MBU}{multi-bit upset}
\newacronym{let}{LET}{linear energy transfer}
\newacronym{sdc}{SDC}{silent data corruption}
\newacronym{due}{DUE}{detectable unrecoverable error}
\newacronym{tid}{TID}{total ionizing dose}
\newacronym{rhbd}{RHBD}{radiation hardening by design}
\newacronym{ge}{GE}{gate equivalent}
\newacronym{udma}{$\mu$DMA}{I/O DMA}
\newacronym{gpio}{GPIO}{}
\newacronym{dma}{DMA}{direct memory access}
\newacronym{tcdm}{TCDM}{tightly coupled data memory}
\newacronym{pdk}{PDK}{process design kit}
\newacronym{pcb}{PCB}{printed circuit board}
\newacronym{sram}{SRAM}{static random-access memory}
\newacronym{tmr}{TMR}{triple modular redundancy}
\newacronym{dmr}{DMR}{dual modular redundancy}
\newacronym{secded}{SECDED}{single error correction, double error detection}
\newacronym{mftf}{MFTF}{mean fluence to failure}
\newacronym{mttf}{MTTF}{mean time to failure}
\newacronym{rpi}{RPi}{Raspberry Pi}
\newacronym{ff}{FF}{flip-flop}
\newacronym{fpga}{FPGA}{field-programmable gate array}
\newacronym{fec}{FEC}{forwards error correction}
\newacronym{obi}{OBI}{Open Bus Interface}
\newacronym{axi}{AXI5}{Advanced eXtensible Interface 5}
\newacronym{apb}{APB5}{Advanced Peripheral Bus 5}
\newacronym[longplural=networks-on-chip]{noc}{NoC}{network-on-chip}
\newacronym{fo4}{FO4}{fan-out of 4 delay}
\newacronym{tmrg}{TMRG}{Triple Modular Redundancy Generator}
\newacronym{rtl}{RTL}{register transfer level}
\newacronym{uart}{UART}{}
\newacronym{jtag}{JTAG}{}
\newacronym{io}{IO}{input/output}
\newacronym{fsm}{FSM}{finite state machine}
\newacronym{sota}{SotA}{state-of-the-art}
\newacronym{relOBI}{relOBI}{reliable OBI}
\newacronym{lsu}{LSU}{load-store unit}
\newacronym{mcu}{MCU}{microcontroller unit}
\newacronym{cmos}{CMOS}{}

\definecolor{PULPRed}{HTML}{A8322C}
\definecolor{PULPBlue}{HTML}{1269B0}
\definecolor{PULPGreen}{HTML}{168638}
\definecolor{PULPOrange}{HTML}{F29545}
\definecolor{PULPPurple}{HTML}{910569}
\definecolor{PULPOlive}{HTML}{48592C}
\definecolor{PULPMarine}{HTML}{007996}
\definecolor{PULPGray}{HTML}{ABABAB}
\definecolor{Red}{HTML}{FF0000}

\def\BibTeX{{\rm B\kern-.05em{\sc i\kern-.025em b}\kern-.08em
    T\kern-.1667em\lower.7ex\hbox{E}\kern-.125emX}}
\begin{document}

\title{Who Checks the Checker?\\Enhancing Component-level Architectural SEU Fault Tolerance for End-to-End SoC Protection
\thanks{This work has received funding from the Swiss State Secretariat for Education, Research, and Innovation (SERI) under the SwissChips initiative.}
}

\author{
\IEEEauthorblockN{Michael Rogenmoser\IEEEauthorrefmark{1}\orcidlink{0000-0003-4622-4862},
Philippe Sauter\IEEEauthorrefmark{1}\orcidlink{0009-0001-6504-8086},
Chen Wu\IEEEauthorrefmark{1}\orcidlink{0009-0006-5417-2870},
Angelo Garofalo\IEEEauthorrefmark{1}\IEEEauthorrefmark{2}\orcidlink{0000-0002-7495-6895},
Luca Benini\IEEEauthorrefmark{1}\IEEEauthorrefmark{2}\orcidlink{0000-0001-8068-3806}}
\IEEEauthorblockA{\IEEEauthorrefmark{1}\textit{ETH Zürich}, Zürich, Switzerland}
\IEEEauthorblockA{\vspace{-10pt}\IEEEauthorrefmark{2}\textit{University of Bologna}, Bologna, Italy}
}
\maketitle

\begin{abstract}
\cGls{seu} fault tolerance for \cglspl{soc} in radiation-heavy environments is often addressed by architectural fault-tolerance approaches protecting individual \cgls{soc} components (e.g., cores, memories) in isolation. However, the protection of voting logic and interconnections among components is also critical, as these become single points of failure in the design. We investigate combining multiple fault-tolerance approaches targeting individual \cgls{soc} components, including interconnect and voting logic to ensure end-to-end \cgls{soc}-level architectural \cgls{seu} fault tolerance, while minimizing implementation area overheads. Enforcing an overlap between the protection methods ensures hardening of the whole design without gaps, while curtailing overheads. We demonstrate our approach on a RISC-V microcontroller \cgls{soc}. \cgls{seu} fault-tolerance is assessed with simulation-based fault injection. Overheads are assessed with full physical implementation. Tolerance to over \SI{99.9}{\percent} of faults in both RTL and implemented netlist is demonstrated. Furthermore, the design exhibits \SI{22}{\percent} lower implementation overhead compared to a single global fault-tolerance method, such as fine-grained triplication.
\end{abstract}

\glsresetall

\section{Introduction}

Fault tolerance of integrated circuits and \cglspl{soc} is of crucial importance for systems operating in radiation-heavy or safety-critical environments. Space~\cite{shin_use_2016}, high-energy physics detectors~\cite{andorno_radiation-tolerant_2023}, avionics~\cite{faubladier_safety_2008}, and functional-safety automotive systems~\cite{ballan_evaluation_2020} increasingly deploy complex safe microcontroller-class \cglspl{soc} that must continue operating correctly even under \cglspl{seu} and \cglspl{set}. In many of these domains, even a single undetected or uncorrected error is unacceptable, as silent data corruption can compromise mission objectives, safety guarantees, or system-wide dependability requirements. On the other hand, the area overhead of fault tolerance is always a concern as it implies a significant cost increase.

Commercial designs targeting space applications, such as the NOEL-V \cgls{soc}~\cite{andersson_gaisler_2022}, often rely on radiation-hardened technologies to achieve fault-tolerance, accompanied by architectural solutions such as \cgls{ecc}. While these technologies can reduce the likelihood of \cglspl{seu} in the design, they do not prevent them completely. Further, they add area, timing, and power overhead, and are not always available, especially for modern technology nodes.

As an alternative, fault-tolerance can be addressed at the architectural level. Fine-grained triplication of the entire \cgls{soc} design~\cite{andorno_radiation-tolerant_2023, walsemann_strv_2023} is a classic approach in high-energy physics, leveraging tools such as \cgls{tmrg}~\cite{kulis_single_2017}, which triplicates the entire design and adds voters throughout. While these solutions avoid depending on radiation-hardened technologies, they come at a very high cost,  showing a \SIrange{4.8}{6.9}{\times} area increase~\cite{andorno_radiation-tolerant_2023} compared to the baseline, unprotected, \cgls{soc}.

To reduce the area overhead, numerous architectural fault-tolerance approaches have been proposed to protect individual \cgls{soc} components (e.g., cores, memories) with component-specific approaches. A common technique is lockstepped processor cores execution~\cite{iturbe_arm_2019, sim_dual_2020, sarraseca_safels_2023, mach_lockstep_2024, rogenmoser_hybrid_2025}, which enables error detection at the cores' boundary through checking or voting mechanisms. Other methods~\cite{shukla_efficient_2023, santos_enhancing_2023} protect critical units within the processor cores, in an effort to further decrease overhead with respect to full-core lockstep. Additional work~\cite{ulbricht_tetrisc_2023} adds \cgls{tmr} to \cglspl{ff} in the design, along with dedicated lockstepping of the processor and fault monitors, but still fails to address \cglspl{set} in combinational logic connecting the \cgls{tmr} \cglspl{ff}.
Yet all these approaches introduce a new vulnerability: the voters and checking logic themselves, which remain unprotected. While small in area, these components become critical points of failure. To address this issue, the voters may require an implementation based on radiation-hardened gates and libraries~\cite{oliveira_fault_2022}. Unfortunately, such gates and libraries are not always available, especially when targeting advanced technologies.

Furthermore, the combination of component-level architectural fault tolerance techniques is rarely examined in terms of how they interact when applied together to protect an entire \cgls{soc}. Prior work~\cite{rogenmoser_trikarenos_2025} evaluating a \cgls{soc} protected with \cgls{ecc} memory and \cgls{tcls} processor cores clearly shows that, especially in extreme environments, only applying such techniques is insufficient, as the remaining vulnerable components at the boundaries and outside the protected regions have a significant residual impact on end-to-end \cgls{soc}-level reliability. The necessity to properly protect peripherals is also highlighted in~\cite{de_oliveira_evaluating_2020}, yet it is often left as a minor concern. These observations strongly motivate the need for end-to-end methods and their evaluation that ensure that all components, including voting, encoding, and interconnect logic, are protected while minimizing area overheads.

To address these challenges, we propose targeted protection methods for each subsystem in a \cgls{soc}, with a novel \textit{overlapping protection approach}. Overlap, where the checking logic of one domain using method A is included within the connected protection domain using method B, enforces that the boundary logic between protection domains, including voters and \cgls{ecc} decoders/encoders, is never left exposed, ensuring that errors in the checking logic itself are detected and corrected.  We demonstrate our approach on \texttt{croc}, a compact RISC-V \cgls{mcu} \cgls{soc}, and show through simulation-based fault injection that overlapping protection defines a new Pareto-optimal point in the reliability–area design space, achieving over \SI{99.9}{\percent} fault coverage while reducing area by \SI{22}{\percent} compared to full triplication.

The contributions of the paper are:
\begin{itemize}
    \item A novel architectural fault-tolerance overlapping method to ensure detection and correction even for the checkers and voters at the boundary of protected components (without resorting to rad-hard gates).
    \item A complete fault-tolerant RISC-V-based \cgls{mcu} \cgls{soc} using dedicated architectural fault tolerance methods for each component, and exploiting overlapping protection, to achieve full end-to-end fault-tolerance.
    \item Analysis of the timing and area impacts of incrementally adding protection methods.
    \item A comprehensive validation of the proposed architecture demonstrating tolerance to faults throughout the design and in typically vulnerable voters, using both \cgls{rtl} and synthesized netlist for fault injection simulation.
\end{itemize}

\section{Background}

\subsection{The \emph{\texttt{croc}} Microcontroller unit}\label{sec:croc}

The \texttt{croc} \cgls{mcu} design~\cite{sauter_croc_2025} is a simple, open-source\footnote{\label{url:croc}\url{https://github.com/pulp-platform/croc}} \cgls{soc}. As illustrated in \cref{fig:refarch}, it features a single 32-bit 2-stage RISC-V core with separate instruction fetch and data interfaces supporting the \cgls{obi} protocol. Connected through an \cgls{obi} interconnect, two \cgls{sram} banks offer separated data and instruction memory for a total of \SI{16}{\kibi\byte} of memory. The \cgls{soc} also features a timer module, as well as \cgls{uart} and \cgls{gpio} peripherals for \cgls{io}. Finally, to enable proper programming and basic functionality, the \cgls{soc} exposes a \cgls{jtag} interface with a RISC-V debug module, as well as some dedicated control registers for programming and boot.

\subsection{Fault \& Protection Model: Single Events are Corrected}

In this paper, we focus on all transient single faults (\cglspl{seu} \& \cglspl{set}) and the accumulation thereof~\cite{dodd_basic_2003}. Observing at a logic-level, the primary expected effects are \cglspl{seu}, bit-flips in state-saving elements such as \cglspl{ff} and \cgls{sram} bit-cells. The other expected effects are \cglspl{set}, transient bit-flips in combinational logic, which are less likely to affect the behavior of the design. However, as clock frequencies increase in modern technologies, these are more likely to be captured by \cglspl{ff}~\cite{di_mascio_open-source_2021, walsemann_fault_2024}. While simultaneous faults are not considered directly, tested \cglspl{set} may induce multiple faulty \cglspl{seu}, and \cglspl{seu}, if not corrected, can accumulate. Furthermore, the design aim is to correct errors and continue operation where possible rather than simply detect faults.

\section{Architecture}

\subsection{Fault-tolerant \emph{\texttt{croc}} Architecture}

\subsubsection{ECC memory}
The first adaptation of the \texttt{croc} \cgls{mcu}, described in \cref{sec:croc}, focuses on the on-chip \cgls{sram}. \cGls{ecc} is ideally suited to ensure single faults are correctable without a large area penalty. As such, we ensure each 32-bit word is encoded with 7 additional bits for \cgls{secded} using a Hsiao code~\cite{hsiao_class_1970}. Each \cgls{sram} bank features its own \cgls{ecc} encoder and decoder, with additional logic to support byte-wise writes in a read-modify-write fashion, immediately responding and delaying subsequent accesses for limited performance impact. Furthermore, to ensure errors do not accumulate within the \cgls{sram}, a scrubber is added to the design. With a configurable period, it reads all data words in the \cgls{sram}, correcting any erroneous words. To avoid impacting performance, it defers its access in case there is an external access to the memory bank. Furthermore, the scrubber also directly corrects any erroneous reads it monitors from external actors.

\begin{figure}[t]%
\centering%
\subfloat[\label{fig:refarch}Baseline \texttt{croc}\\architecture.]{%
\includegraphics[width=0.37\linewidth,valign=c]{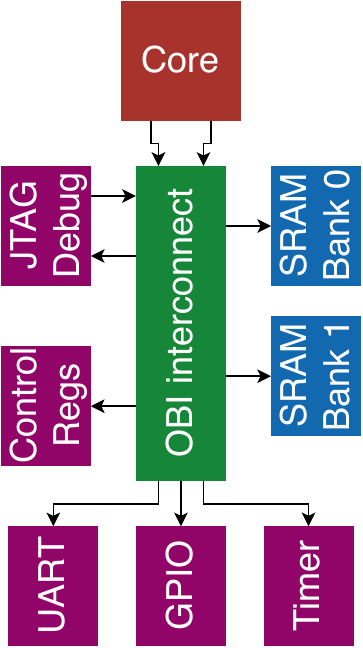}%
}%
\hfill%
\subfloat[\label{fig:relarch}Reliable \texttt{croc} architecture.]{%
\includegraphics[width=0.53\linewidth,valign=c]{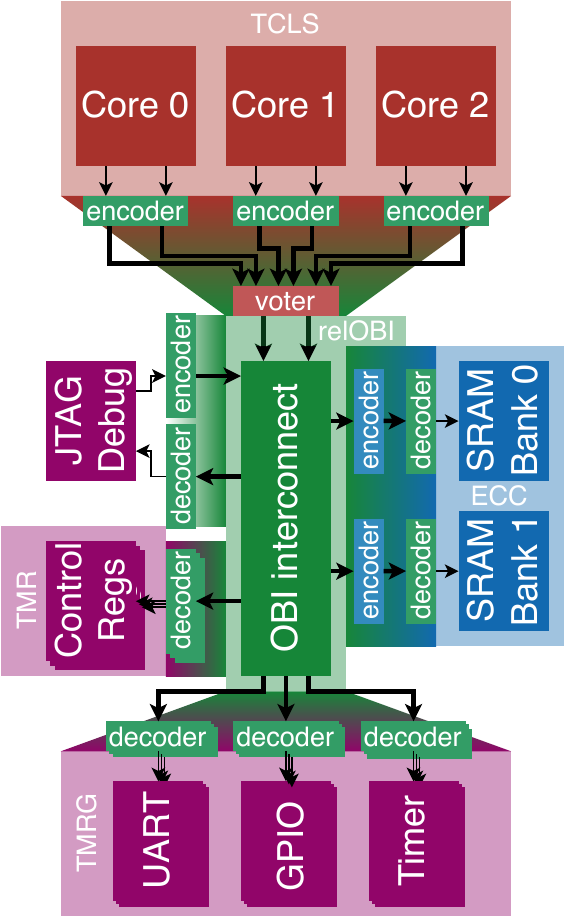}%
}%
\caption{Block diagram of the \texttt{croc} architecture, with (b) showing protected domains highlighted.}%
\label{fig:arch}%
\vspace{-12pt}%
\end{figure}

\subsubsection{TCLS core}
The second adaptation focuses on the processor core. Leveraging the work in~\cite{rogenmoser_hybrid_2025}, a static \cgls{tcls} configuration is implemented for the RISC-V core. The entire processor core is triplicated, with identical inputs connected to the three cores, and voters added for the cores' output bits, most of which are \cgls{obi} bus signals. In case an error is detected, a software resynchronization routine is executed in an interrupt service routine to ensure any errors remaining within the cores' registers are corrected.

\subsubsection{relOBI interconnect}
The third adaptation focuses on the \cgls{obi} interconnect between the bus's managers, mainly the processor core, and subordinates, including the \cgls{sram} banks and the peripherals. \cGls{relOBI}~\cite{rogenmoser_relobi_2025} augments the \cgls{obi} protocol with fault tolerance, where critical handshake signals, as well as the interconnection modules' internals, are protected with \cgls{tmr}, ensuring consistency of the transfers and the interconnect state. Data, address, and additional information passed through the interconnect are protected with \cgls{ecc} to avoid excessive area overhead.

\subsubsection{TMR peripherals}
The fourth adaptation applies \cgls{tmr} to many of the remaining components in the design. This includes the timer, \cgls{uart}, and \cgls{gpio} peripherals, which are fully protected using \cgls{tmrg}~\cite{kulis_single_2017}, where we ensure triplicated voters are added after every register to correct any internal latent faults. Furthermore, any memory-mapped registers, such as the \cgls{soc} control registers, and the corresponding signals and \cglspl{fsm}, such as the \cgls{ecc} scrubbing unit, are also triplicated. \Cref{fig:arch} shows all modifications added to the design.

\subsubsection{Fault monitoring}
To ensure traceability, faults are monitored and collected in fault counter registers. This enables verification of protection mechanisms both in software and after tapeout. To improve timing, the collected fault signals are pipelined prior to incrementing the counter. Access is ensured to be reliable, but the counter values are unprotected, as they are not part of the operational design.

\subsubsection{Remaining components}
Some components in the design are not protected. The debug module, required for bringup and testing, is not designed to be used in an active environment and is thus not hardened. To ensure it does not affect the system, its signals and bus interfaces can be isolated with dedicated units that are protected.

\subsection{Overlapping protection}

\begin{figure}[t]%
\centering%
\subfloat[\label{fig:overlap_a}Non-overlapped protection.]{%
\includegraphics[width=\linewidth]{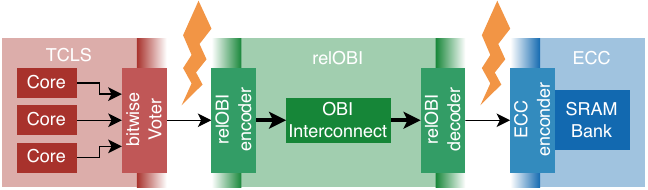}%
}%
\\%
\subfloat[\label{fig:overlap_b}Overlapped protection.]{%
\includegraphics[width=\linewidth]{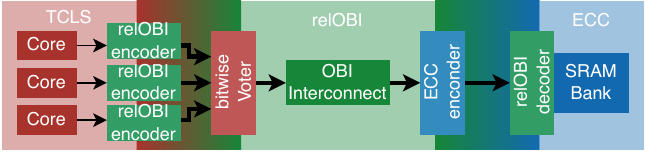}%
}%
\caption{Section of the design in~\cref{fig:relarch}. Overlapping fault tolerance methods remove vulnerabilities in connecting signals and voting, encoding, and decoding logic in gaps between protected domains.}%
\label{fig:overlapping}%
\vspace{-12pt}%
\end{figure}

Naively combining different fault tolerance methods would result in an implementation where each protection method is considered individually. Investigating a slice of the protected design, \cref{fig:overlap_a} shows a processor core protected with \cgls{tcls} and an \cgls{ecc}-protected \cgls{sram} memory. Inserting a \cgls{relOBI}-protected interconnect in between still leaves parts of this system, the connection between protected regions, vulnerable to faults. Not only are the connections vulnerable, but the voting, encoding, and decoding logic also remain points of failure.

Thus, when combining \cgls{ecc}-protected \cgls{sram} and \cgls{tcls} processor cores with \cgls{relOBI}, we ensure an overlap of these protection methods to remove any gaps in the protection, as shown in \cref{fig:overlap_b}. Instead of voting on raw data, the \cgls{relOBI} encoders and decoders are integrated within the lockstepped domain, performing the \cgls{tcls} voting on encoded signals.
This ensures that errors in the voters for \cgls{tcls} are detected and corrected as errors of the \cgls{relOBI} interface, and that errors in the \cgls{relOBI} encoding and decoding are detected as \cgls{tcls} faults, as the units are triplicated with each lockstepped core. Furthermore, as the data encoding in \cgls{sram} is identical to the encoding within the \cgls{relOBI} bus signals, it is directly reused when storing the data to memory, removing the need for additional \cgls{ecc} encoding and decoding at the \cgls{sram} bank. Therefore, while any data error in \cgls{sram} may be propagated to the \cgls{relOBI} interconnect, the corresponding decoders will still correct the fault data word, and vice versa.

\section{Experimental Setup}

To investigate both implementation overheads and fault tolerance benefits, different configurations for the \texttt{croc} \cgls{soc} are investigated, each with incrementally more fault detection and correction methods as outlined above.
\begin{enumerate}[label=Cfg.~\arabic*:, left=10pt]
    \setcounter{enumi}{-1}
    \item No protection / reference
    \item \cgls{ecc}-protected memory
    \item Protected memory \& \cgls{tcls} cores
    \item Protected memory, cores \& \cgls{relOBI} interconnect (with overlap)
    \item Full protection (with overlap)
    \item[\cgls{tmrg}:] Related work, explained in \cref{sec:related}
\end{enumerate}

\subsection{Implementation Flow}

Following the existing setup for the \texttt{croc} \cgls{soc}\footref{url:croc}, the different configurations for \texttt{croc} were synthesized with \textit{Yosys} and placed and routed with \textit{OpenROAD} using the open-source IHP \SI{130}{\nano\meter} \cgls{pdk} with sg13g2 standard cells at typical conditions. To ensure triplicated elements in the design were not optimized away, hierarchies for these triplicated components or the required fanout and voter cells were kept, ensuring their boundaries and signals are not optimized away. Simple cross-boundary optimizations, such as constant propagation and removal of undriven signals, were allowed, as this does not affect the integrity of the \cgls{tmr} protection.

\subsection{Fault Analysis}

To analyze the \texttt{croc} system under faults, we utilize \textit{Synopsys\,\textsuperscript{\tiny\textregistered} VC Z01X\,\textsuperscript{\tiny\texttrademark} X-2025.06}, a commercial concurrent fault simulator. Following an initial reference simulation of the \texttt{croc} \cgls{soc} running a particular workload, it concurrently simulates the same design, injecting randomly sampled single faults, tracking the outcome for each simulation.

\begin{figure*}[t]%
    \subfloat[\label{fig:die0}Cfg.~0: Unprotected \\\quad]{%
        \includegraphics[width=0.1952\linewidth]{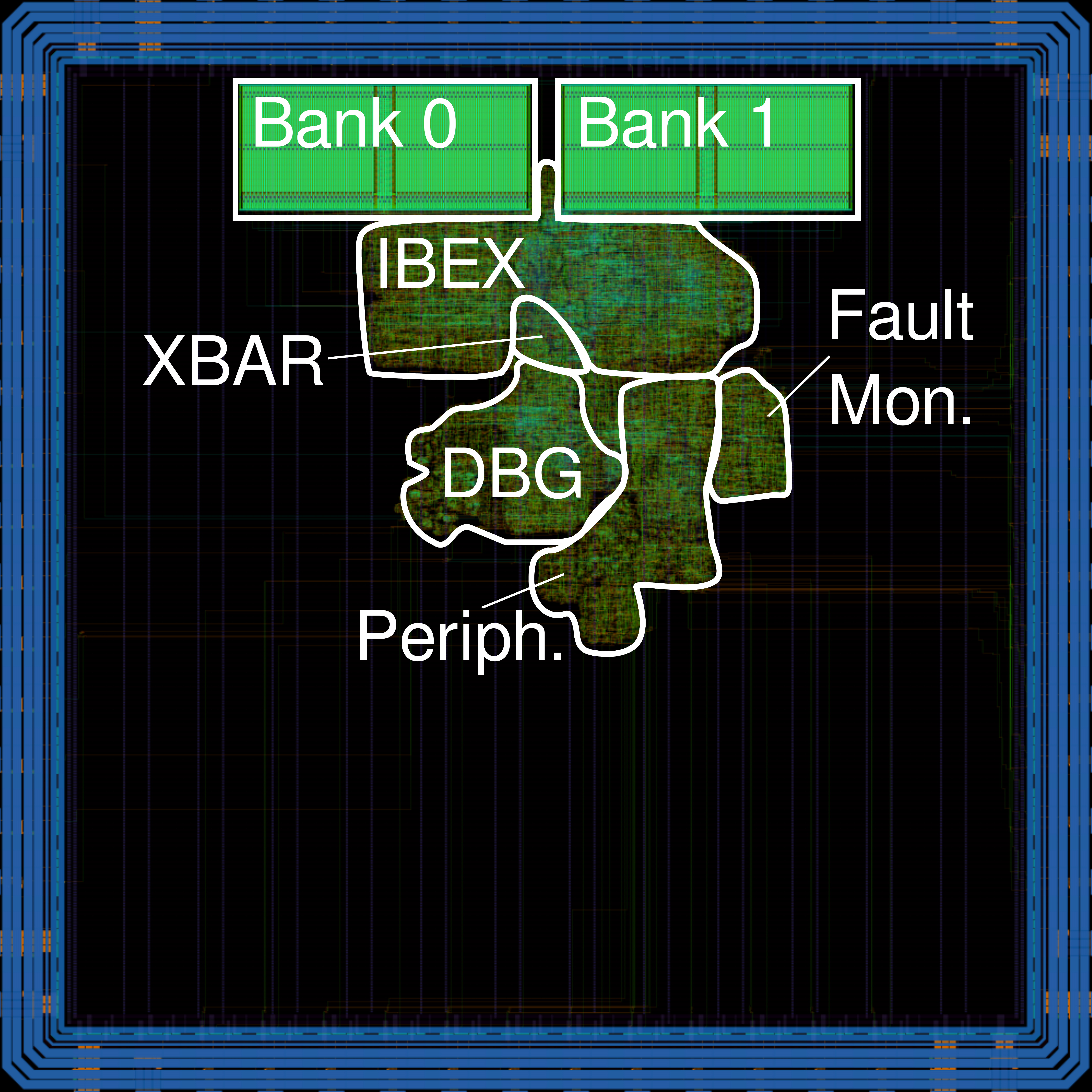}%
    }%
    \hfill%
    \subfloat[\label{fig:die1}Cfg.~1: \cgls{ecc} Memory \\\quad]{%
        \includegraphics[width=0.1952\linewidth]{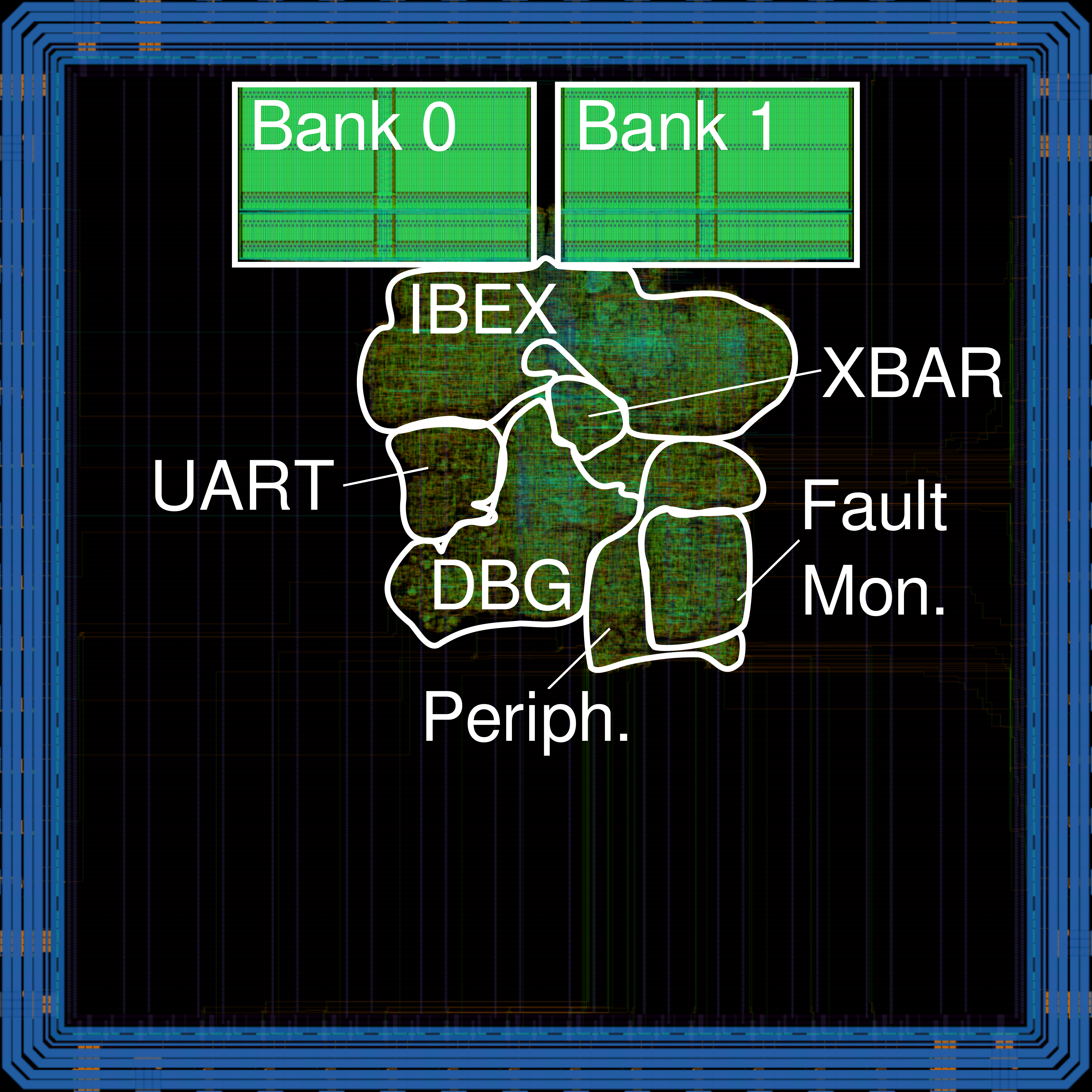}%
    }%
    \hfill%
    \subfloat[\label{fig:die2}Cfg.~2: \cgls{ecc} Memory,\\\cgls{tcls} cores]{%
        \includegraphics[width=0.1952\linewidth]{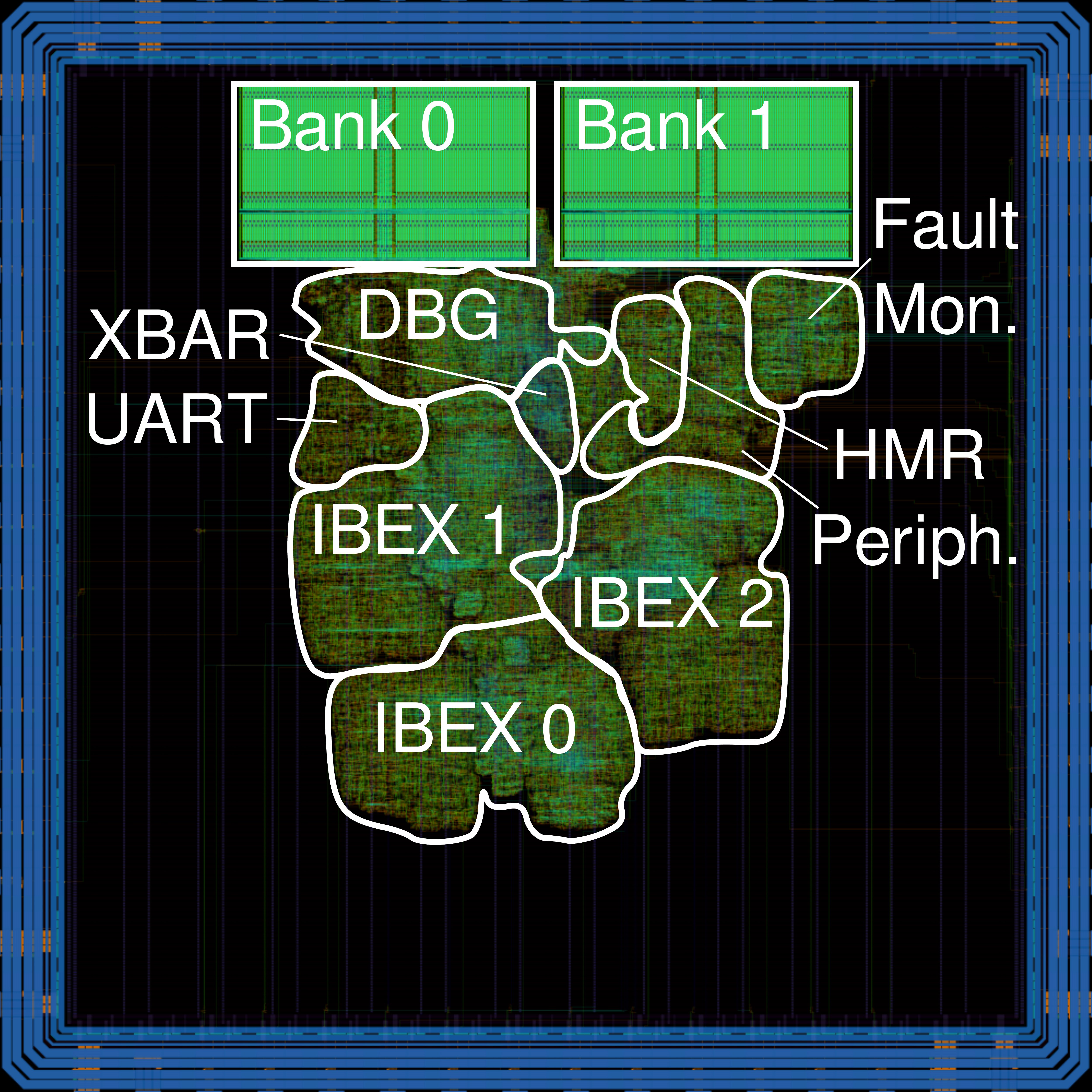}%
    }%
    \hfill%
    \subfloat[\label{fig:die3}Cfg.~3: \cgls{ecc}, \cgls{tcls}, \cgls{relOBI} interconnect]{%
        \includegraphics[width=0.1952\linewidth]{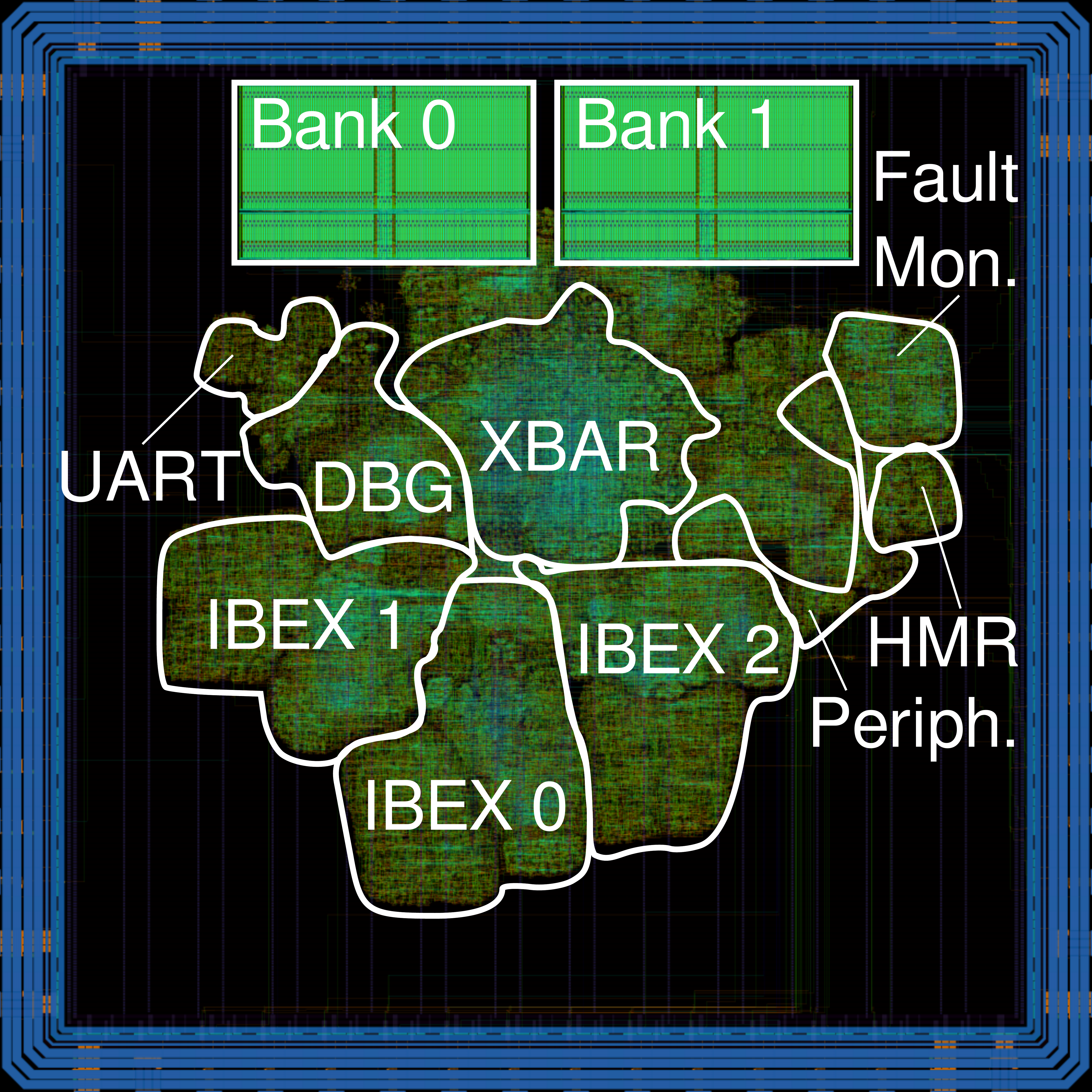}%
    }%
    \hfill%
    \subfloat[\label{fig:die4}Cfg.~4: Full protection\\\quad]{%
        \includegraphics[width=0.1952\linewidth]{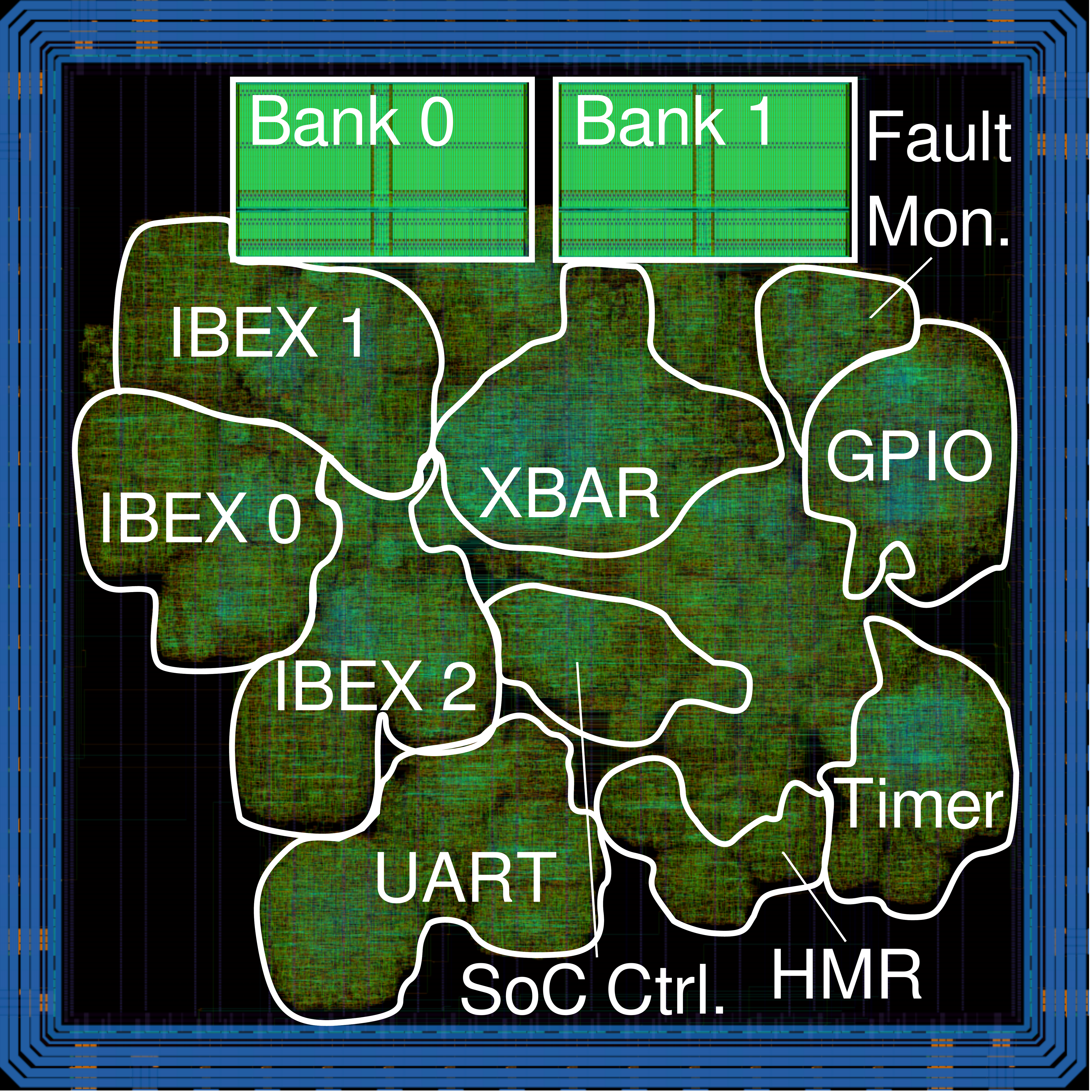}%
    }%
    \caption{\label{fig:dieshots}Annotated implementation layout render for all tested configurations}%
    \vspace{-12pt}%
\end{figure*}

\subsubsection{Workload}
Prior to injecting a fault, the simulation preloads the application into the on-chip \cgls{sram} and starts the processor at the corresponding binary location. Thereafter, the core clears the remaining \cgls{sram} to ensure there is no faulty data that could indicate an \cgls{ecc} fault. After clearing the fault monitor, the scrubber is started, thereafter allowing the core to enter the main application. This part of the simulation is excluded from fault injection to ensure a reproducible setup, and as this setup process is not expected as part of an operational application loop.

For proper analysis, the tested software is designed to exercise most parts of the \cgls{soc}, emulating a single loop of a longer application. Thus, the application includes a simple output string transferred over the \cgls{uart} module and toggling of the \cgls{gpio} pins to verify the \cgls{io} pins are behaving as expected. A short timer delay is added to ensure the core's interrupts, as well as the timer, are functioning as expected. Finally, a single CoreMark\textsuperscript{\tiny\textregistered} iteration, along with its corresponding check, is executed to ensure the processor performs as expected, storing the number of errors along with a status bit into a return value register to end the application loop.

\subsubsection{Monitoring}
To detect the outcome of individual simulations, \textit{VC Z01X\,\textsuperscript{\tiny\texttrademark}} allows for comparison of individual signals to their values in a reference, non-faulty simulation. By default, we assume a correct outcome with no detected erroneous behavior, which means the injected fault was masked. The \cgls{soc}'s output pins, the return value register containing the number of errors in the CoreMark\textsuperscript{\tiny\textregistered} calculation, and the core's exception indicator are compared every cycle. This allows for the detection of any functional issues in the tested \cgls{soc} components, detected as data errors, processor core exceptions, and timeouts or early terminations.

While some faults may affect the tested iteration loop of the applications, some faults may only become apparent in the following loop iterations or longer-duration applications. As such, at the end of the simulation, we compare the data stored in \cgls{sram} to the reference simulation to detect latent faults. These latent faults may be due to a \cgls{seu} directly in the \cgls{sram} that was never overwritten, or due to a corrupted internal state from fault processing that was not otherwise exposed.

Finally, for configurations that contain fault tolerance mechanisms, any correction is logged and differentiated from a simple masked fault outcome. Furthermore, uncorrectable faults, such as multi-bit faults in a single \cgls{ecc} word, are detected and logged as such, irrespective of their correct or incorrect outcome, as we expect a direct triggering of external reset or watchdog logic.

\subsubsection{Fault Injection}
For a proper understanding of the behavior of the system under different fault scenarios, 3 sets of experiments are performed, always injecting \SI{100000}{faults} for each configuration. The first two experiments emulate \cgls{seu} faults directly by injecting \texttt{FF} faults into the RTL simulation, flipping the value of a \cgls{ff} or \cgls{sram} cell for a single cycle and thereafter allowing the value to be overwritten, much like a \cgls{seu} affecting a register. While the first scenario targets all possible state bits throughout the \cgls{soc}, randomly selecting a signal and time point within the active application with equal probability, the second scenario excludes the \cgls{sram} from being injected. As the \SI{16}{\kilo\byte} accounts for over \SI{90}{\percent} of injectable bits in the design, excluding these allows for more extensive analysis and can help account for differing \cgls{ff} and \cgls{sram} cell vulnerabilities.

The third scenario targets the simulation of \cglspl{set} using the synthesized netlist. To simulate this, we inject \texttt{PRIM} faults, bit-flips in the output of Verilog primitives used to describe the function of the cells in the design, keeping the fault applied across a clock edge to ensure the fault effect propagates. While this may not model \cglspl{set} representatively, it ensures the tested faults have a chance to affect the design. As the \cgls{sram} is simulated as a behavioral model, it is also excluded from these simulations.

\section{Results}

\subsection{Implementation}
All tested \texttt{croc} configurations were synthesized and subsequently implemented on an identical 3$\times$\SI{3}{\milli\meter\squared} die at a target frequency of \SI{60}{\mega\hertz} to compare the required area. The placed and routed designs are shown in \cref{fig:dieshots} for all configurations, with more detailed results shown in \cref{tab:implementation}.

\begin{table}[h]
    \centering
    \caption{Post-layout implementation results for \texttt{croc} Cfgs.}
    \begin{tabular}{@{}rHc@{ / }cS[table-format=3, table-space-text-post={\,\si{\mega\hertz}}]<{\,\si{\mega\hertz}}S[table-format=5]c@{}}\toprule
        \multirow{2}{*}{Cfg.} & \multirow{2}{*}{Description}             & \multicolumn{2}{@{}c}{Used Core Area} & \multicolumn{1}{c}{Max. achieved}   & \multicolumn{1}{c}{\multirow{2}{*}{Registers}} & Area     \\
                              &                                          & \multicolumn{2}{@{}c}{@ \SI{60}{\mega\hertz}} & \multicolumn{1}{c}{Frequency}       & & Overhead \\\midrule
        0                     & No protection                            & \SI{1.057}{\milli\meter\squared} & \SI{146}{\kilo GE} & 112 & 5113 & \SI{1.00}{\times} \\
        1                     & \cgls{ecc} memory                        & \SI{1.286}{\milli\meter\squared} & \SI{177}{\kilo GE} & 108 & 5254 & \SI{1.22}{\times} \\
        2                     & \cgls{ecc} memory \& \cgls{tcls}         & \SI{1.758}{\milli\meter\squared} & \SI{242}{\kilo GE} & 95 & 9358 & \SI{1.66}{\times} \\
        3                     & \cgls{ecc} memory, \cgls{tcls} \& \cgls{relOBI} & \SI{2.171}{\milli\meter\squared} & \SI{299}{\kilo GE} & 65 & 10175 & \SI{2.05}{\times} \\
        4                     & Full protection                          & \SI{2.867}{\milli\meter\squared} & \SI{395}{\kilo GE} & 62 & 13799 & \SI{2.71}{\times} \\
        \textit{\cgls{tmrg}} & \textit{\cgls{tmrg}} & \SI{3.676}{\milli\meter\squared} & \SI{507}{\kilo GE} & 70 & 11964 & \SI{3.48}{\times} \\
        \bottomrule
    \end{tabular}
    \label{tab:implementation}
\end{table}

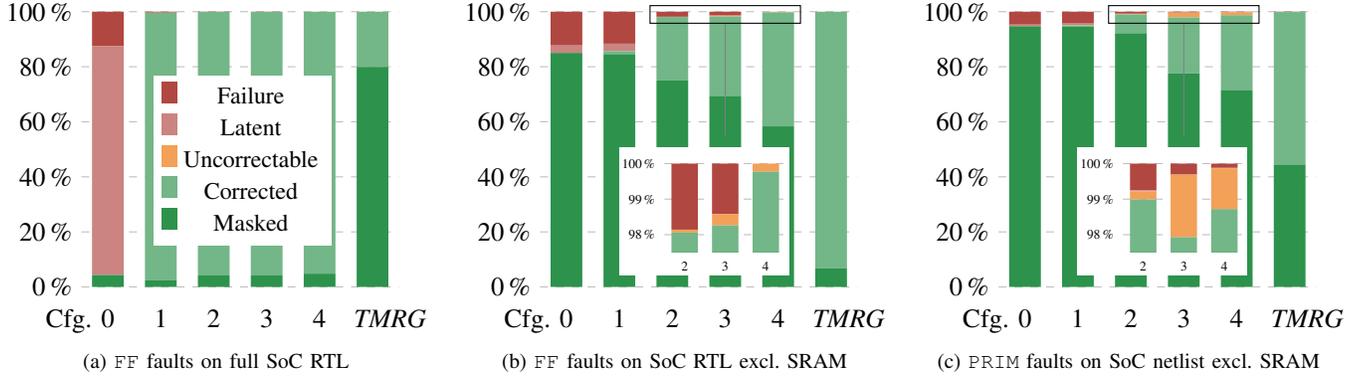
\begin{figure*}[t]%
    \centering%
    \subfloat[\label{fig:measurement1}\texttt{FF} faults on full \cgls{soc} \cgls{rtl}]{
\begin{tikzpicture}
    \begin{axis}[
        ybar stacked,
        ymin=0,
        ymax=107,
        xtick={0,1,2,3,4,5},
        xticklabels={{Cfg. 0\hphantom{Cfg. }},{1},{2},{3},{4},{\textit{\hphantom{RG}TMRG}}},
        legend style={at={(axis cs:0.85, 15)},anchor=south west, draw=none, font=\small},
        reverse legend=true,
        ytick={0,20,40,60,80,100},
        yticklabels={{\SI{0}{\percent}}, \SI{20}{\percent}, \SI{40}{\percent}, \SI{60}{\percent}, \SI{80}{\percent}, \SI{100}{\percent}},
        bar width=12pt,
        ymajorgrids=true,
        grid style=dashed,
        xtick style={draw=none},
        ytick style={draw=none},
        xticklabel shift = 5pt,
        width=0.32\textwidth,
        height=5.5cm,
        axis line style={draw=none},
    ]
        \addplot [fill=PULPGreen!90, draw opacity=0] coordinates {(0,4.34) (1,2.49) (2,4.17) (3,4.16) (4,4.83) (5,79.94)};
        \addplot [fill=PULPGreen!60, draw opacity=0] coordinates {(0,0) (1,97.08) (2,95.73) (3,95.73) (4,95.15) (5,20.06)};
        \addplot [fill=PULPOrange!90, draw opacity=0] coordinates {(0,0) (1,0) (2,0.004) (3,0.014) (4,0.020) (5,0)};
        \addplot [fill=PULPRed!60, draw opacity=0] coordinates {(0,83.12) (1,0.08) (2,0) (3,0) (4,0) (5,0)};
        \addplot [fill=PULPRed!90, draw opacity=0] coordinates {(0,12.54) (1,0.35) (2,0.10) (3,0.09) (4,0) (5,0)};
        \legend{Masked,Corrected,Uncorrectable,Latent,Failure};
    \end{axis}
\end{tikzpicture}
}%
    \hfill%
    \subfloat[\label{fig:measurement2}\texttt{FF} faults on \cgls{soc} \cgls{rtl} excl. \cgls{sram}]{
\begin{tikzpicture}[spy using outlines={}]
    \begin{axis}[
        ybar stacked,
        ymin=0,
        ymax=107,
        xtick={0,1,2,3,4,5},
        xticklabels={{Cfg. 0\hphantom{Cfg. }},{1},{2},{3},{4},{\textit{\hphantom{RG}TMRG}}},
        legend style={at={(axis cs:-0.1, 3)},anchor=south west, draw=none},
        reverse legend=true,
        ytick={0,20,40,60,80,100},
        yticklabels={{\SI{0}{\percent}}, \SI{20}{\percent}, \SI{40}{\percent}, \SI{60}{\percent}, \SI{80}{\percent}, \SI{100}{\percent}},
        bar width=12pt,
        ymajorgrids=true,
        grid style=dashed,
        xtick style={draw=none},
        ytick style={draw=none},
        xticklabel shift = 5pt,
        width=0.32\textwidth,
        height=5.5cm,
        axis line style={draw=none},
    ]
        \addplot [fill=PULPGreen!90, draw opacity=0] coordinates {(0,85.02) (1,84.57) (2,75.00) (3,69.39) (4,58.34) (5,6.80)};
        \addplot [fill=PULPGreen!60, draw opacity=0] coordinates {(0,0) (1,1.11) (2,23.06) (3,28.87) (4,41.43) (5,93.20)};
        \addplot [fill=PULPOrange!90, draw opacity=0] coordinates {(0,0) (1,0) (2,0.08) (3,0.32) (4,0.23) (5,0)};
        \addplot [fill=PULPRed!60, draw opacity=0] coordinates {(0,2.70) (1,2.69) (2,0) (3,0) (4,0) (5,0)};
        \addplot [fill=PULPRed!90, draw opacity=0] coordinates {(0,12.28) (1,11.63) (2,1.86) (3,1.42) (4,0) (5,0)};
        \coordinate (spypoint3) at (axis cs:3,99);
    \end{axis}
    \node[pin={[pin distance=1.5cm]270:{%
        \begin{tikzpicture}[
            baseline,
            trim axis left,
            trim axis right,
        ]
            \begin{scope}[on background layer={color=white}]
                \fill (-0.6,-0.3) rectangle (1.65,1.4);
            \end{scope}
            \begin{axis}[
                    ybar stacked,
                    xtick={0,1,2,3,4,5},
                    tiny,
                    xmin=1.5,xmax=4.5,
                    ymin=97.5,ymax=100.5,
                    bar width=10pt,
                    xticklabels={0,1,2,3,4,TMRG},
                    ytick={98,99,100},
                    yticklabels={\SI{98}{\percent}, \SI{99}{\percent},\SI{100}{\percent}},
                    width=3.2cm,
                    height=3cm,
                    ymajorgrids=true,
                    grid style=dashed,
                    xtick style={draw=none},
                    ytick style={draw=none},
                    axis line style={draw=none},
                ]
                \addplot [fill=PULPGreen!90, draw opacity=0] coordinates {(0,85.02) (1,84.57) (2,75.00) (3,69.39) (4,58.34) (5,6.80)};
                \addplot [fill=PULPGreen!60, draw opacity=0] coordinates {(0,0) (1,1.11) (2,23.06) (3,28.87) (4,41.43) (5,93.20)};
                \addplot [fill=PULPOrange!90, draw opacity=0] coordinates {(0,0) (1,0) (2,0.08) (3,0.32) (4,0.23) (5,0)};
                \addplot [fill=PULPRed!60, draw opacity=0] coordinates {(0,2.70) (1,2.69) (2,0) (3,0) (4,0) (5,0)};
                \addplot [fill=PULPRed!90, draw opacity=0] coordinates {(0,12.28) (1,11.63) (2,1.86) (3,1.42) (4,0) (5,0)};
            \end{axis}
        \end{tikzpicture}%
    }},draw,rectangle,minimum height=0.2cm, minimum width=2cm] at (spypoint3) {};
\end{tikzpicture}
}%
    \hfill%
    \subfloat[\label{fig:measurement3}\texttt{PRIM} faults on \cgls{soc} netlist excl. \cgls{sram}]{
\begin{tikzpicture}[spy using outlines={}]
    \begin{axis}[
        ybar stacked,
        ymin=0,
        ymax=107,
        xtick={0,1,2,3,4,5},
        xticklabels={{Cfg. 0\hphantom{Cfg. }},{1},{2},{3},{4},{\textit{\hphantom{RG}TMRG}}},
        legend style={at={(axis cs:0, 10)},anchor=south west, draw=none},
        reverse legend=true,
        ytick={0,20,40,60,80,100},
        yticklabels={\SI{0}{\percent}, \SI{20}{\percent}, \SI{40}{\percent}, \SI{60}{\percent}, \SI{80}{\percent}, \SI{100}{\percent}},
        bar width=12pt,
        ymajorgrids=true,
        grid style=dashed,
        xtick style={draw=none},
        ytick style={draw=none},
        xticklabel shift = 5pt,
        width=0.32\textwidth,
        height=5.5cm,
        axis line style={draw=none},
    ]
        \addplot [fill=PULPGreen!90, draw opacity=0] coordinates {(0,94.71) (1,94.54) (2,92.17) (3,77.57) (4,71.54) (5,44.48)};
        \addplot [fill=PULPGreen!60, draw opacity=0] coordinates {(0,0) (1,0.50) (2,6.82) (3,20.36) (4,27.18) (5,55.40)};
        \addplot [fill=PULPOrange!90, draw opacity=0] coordinates {(0,0) (1,0.14) (2,0.23) (3,1.76) (4,1.15) (5,0)};
        \addplot [fill=PULPRed!60, draw opacity=0] coordinates {(0,0.59) (1,0.51) (2,0.03) (3,0) (4,0.01) (5,0)};
        \addplot [fill=PULPRed!90, draw opacity=0] coordinates {(0,4.70) (1,4.31) (2,0.75) (3,0.31) (4,0.12) (5,0.12)};
        \coordinate (spypoint3) at (axis cs:3,99);
    \end{axis}
    \node[pin={[pin distance=1.5cm]270:{%
        \begin{tikzpicture}[
            baseline,
            trim axis left,
            trim axis right,
        ]
            \begin{scope}[on background layer={color=white}]
                \fill (-0.6,-0.3) rectangle (1.65,1.4);
            \end{scope}
            \begin{axis}[
                    ybar stacked,
                    xtick={0,1,2,3,4,5},
                    tiny,
                    xmin=1.5,xmax=4.5,
                    ymin=97.5,ymax=100.5,
                    bar width=10pt,
                    xticklabels={0,1,2,3,4,TMRG},
                    ytick={98,99,100},
                    yticklabels={\SI{98}{\percent}, \SI{99}{\percent},\SI{100}{\percent}},
                    width=3.2cm,
                    height=3cm,
                    ymajorgrids=true,
                    grid style=dashed,
                    xtick style={draw=none},
                    ytick style={draw=none},
                    axis line style={draw=none},
                ]
                \addplot [fill=PULPGreen!90, draw opacity=0] coordinates {(0,94.71) (1,94.54) (2,92.17) (3,77.57) (4,71.54) (5,44.48)};
                \addplot [fill=PULPGreen!60, draw opacity=0] coordinates {(0,0) (1,0.50) (2,6.82) (3,20.36) (4,27.18) (5,55.40)};
                \addplot [fill=PULPOrange!90, draw opacity=0] coordinates {(0,0) (1,0.14) (2,0.23) (3,1.76) (4,1.15) (5,0)};
                \addplot [fill=PULPRed!60, draw opacity=0] coordinates {(0,0.59) (1,0.51) (2,0.03) (3,0) (4,0.01) (5,0)};
                \addplot [fill=PULPRed!90, draw opacity=0] coordinates {(0,4.70) (1,4.31) (2,0.75) (3,0.31) (4,0.12) (5,0.12)};
            \end{axis}
        \end{tikzpicture}%
    }},draw,rectangle,minimum height=0.2cm, minimum width=2cm] at (spypoint3) {};
\end{tikzpicture}%
}
    \caption{\label{fig:measurement}Fault injection results}%
    \vspace{-12pt}%
\end{figure*}

\textbf{Cfg. 0:} The reference design remains fairly compact, as seen in \cref{fig:die0}, at \SI{1.057}{\milli\meter\squared} of core area used for cells. While the design can achieve up to \SI{112}{\mega\hertz} clock frequency, a typical target is far lower for this low-performance design. The critical path is shown to originate from the controller in the core's ID stage, traverse the core's register file, exit the core through the signals in the instruction fetch interface, negotiate arbitration in the interconnect, and terminate in a register in the core's \cgls{lsu}.

\textbf{Cfg. 1:} Adding \cgls{ecc} to the \cgls{sram} memory banks increases the total \cgls{soc} area utilization by around \SI{1.22}{\times}. While the 7 additional bits for each 32-bit word increase the bits stored by \SI{1.22}{\times}, the \cgls{sram} macros used increase the total utilized \cgls{sram} area from \SI{0.528}{\milli\meter\squared} to \SI{0.714}{\milli\meter\squared}, visible in \cref{fig:die1}. Along with a \SI{0.043}{\milli\meter\squared} overhead for \cgls{ecc} encoding and decoding, the memory scrubber, and the fault monitoring counters, the area impact remains minimal. The addition of encoding logic affects the maximum achieved frequency, with a similar critical path ending in the \cgls{sram} bank after exiting the \cgls{obi} interconnect arbiter and \cgls{ecc} encoding the data word.

\textbf{Cfg. 2:} Triplicating the processor cores using \cgls{tcls} adds 2 additional cores to the design, as well as the required voting and control logic. With each core utilizing \SI{0.211}{\milli\meter\squared}, an additional \SI{0.044}{\milli\meter\squared} completes the \SI{0.472}{\milli\meter\squared} area overhead for the voting logic, configuration registers, and fault monitoring counters, shown in \cref{fig:die2}. The voters and corresponding error detection logic add additional timing overhead to the critical path, along with the increasing complexity of the design, making implementation more challenging.

\textbf{Cfg. 3:} Adding an overlapping \cgls{relOBI} interconnect to the \cgls{soc} adds \SI{0.413}{\milli\meter\squared} to the total area. This is accompanied by an increased implementation complexity shown in \cref{fig:die3} due to the tightly connected triplication of internal modules. Furthermore, due to overlapping, the complexity of other components increases. Triplication also requires more strict module boundary enforcement for implementation, resulting in reduced optimization potential and a more complex design. Furthermore, the additional encoding and subsequent decoding for interconnect signals in \cgls{relOBI}, such as the address, and triplication results in a significant timing impact, increasing the critical path to support at most \SI{65}{\mega\hertz}. Properly pipelining the fetch interface of the design, as is common in higher-performance systems, and improving the \cgls{relOBI} architecture to avoid repeated \cgls{ecc} en-/decoding overhead within a single timing path would alleviate the timing penalty.

\textbf{Cfg. 4:} Applying fine-grained triplication to the \cgls{uart}, \cgls{gpio}, and timer modules, as well as all control registers and corresponding logic, results in an additional \SI{0.696}{\milli\meter\squared} of utilized area. The \cgls{gpio} module increases from \SI{0.028}{\milli\meter\squared} to \SI{0.174}{\milli\meter\squared}, a \SI{6.2}{\times} area increase, due to the large number of registers and the correspondingly required voters after triplication and their fault detection logic. Similarly, the timer module increases from \SI{0.028}{\milli\meter\squared} to \SI{0.150}{\milli\meter\squared}. As the additional components are not on the critical path, only the additional complexity of the design leads to slightly lower maximum frequency.

\subsection{Fault analysis\label{sec:res_fault}}

All fault-free simulations of the different configurations terminated with identical cycle counts, as the fault tolerance mechanisms do not affect performance in the normal case.

\textbf{Cfg. 0:} In the fully unprotected design, \SI{12.54}{\percent} of injections lead to a failure when injecting \texttt{FF} faults on the entire \cgls{soc}. As shown in the first bar of \cref{fig:measurement1}, \SI{83.12}{\percent} of outcomes show correct behavior but still contain latent errors in the \cgls{sram}. With over \SI{90}{\percent} of faults injected into the design being injected into the \cgls{sram}, most of these latent errors simply indicate the flipped bit from injection, but some errors indicate larger impacts on the \cgls{soc} that can lead to a failure of an application further down the line. Excluding faults in \cgls{sram}, we still observe a \SI{12.28}{\percent} fault rate with \SI{2.70}{\percent} of simulations resulting in latent errors, shown in \cref{fig:measurement2}. Results for netlist simulations in \cref{fig:measurement3} are lower, showing a \SI{4.70}{\percent} failure rate and \SI{0.59}{\percent} latent errors. For both scenarios, excluding \cgls{sram}, the latent errors in \cgls{sram} are due to \cgls{soc} errors propagating into \cgls{sram}, as the \cgls{sram} itself is not injected.

\textbf{Cfg. 1:} To address the main source of faults in the complete \cgls{soc}, we add \cgls{ecc} protection to the \cgls{sram} memory banks. This directly addresses all single faults injected into the \cgls{sram}, shown in \cref{fig:measurement1}, as a single error in a word is corrected upon reading, either by the system or the scrubber, resulting in a remaining \SI{0.35}{\percent} failure rate. Errors in the state of the remaining \cgls{soc} still lead to system failures, with  \cref{fig:measurement2,fig:measurement3} showing \SI{11.63}{\percent} of \texttt{FF} faults and \SI{4.31}{\percent} of faults in the netlist leading to a failure, as well as \SI{2.69}{\percent} and \SI{0.51}{\percent} latent errors, respectively. This is similar to the baseline design, as the only protected domain is excluded from these simulations. The slight reduction can be attributed to the additional logic added for encoding and scrubbing, increasing the possible faults, thus reducing the effective fault rate.

\textbf{Cfg. 2:} Tackling the largest active component in the design, the processor core, we observe a \SI{6.2}{\times} and \SI{5.7}{\times} decrease in failure rates for non-memory registers and netlist faults, respectively, to \SI{1.86}{\percent} and \SI{0.75}{\percent}. Studying \cref{fig:measurement2}, we also observe that the number of corrections increases beyond the previous number of failures and latent errors. This is due to both detecting erroneous behavior that may not have led to a failure previously, as well as an increased number of injected faults into the cores, as the random faults are injected evenly distributed across all registers in the \cgls{soc}. With \SI{1.8}{\times} more registers as well as correspondingly more logic now in the processor core due to triplication, faults are also more likely to be injected there.

\textbf{Cfg. 3:} Adding reliability to the interconnect again reduces the failure rate while increasing the correction rate. With this addition, we observe an increase in the number of uncorrectable errors detected, from \SI{0.08}{\percent} and \SI{0.23}{\percent} in Cfg.~2 to \SI{0.32}{\percent} and \SI{1.77}{\percent} in \cref{fig:measurement2,fig:measurement3}, respectively. Uncorrectable errors indicate multiple-bit errors in a single \cgls{ecc}-protected word, and some are traceable to a fault injected in the \cgls{ecc} encoding. Furthermore, faults in the detection or-tree also affect the results.

\textbf{Cfg. 4:} The entire active design is now protected, resulting in no detected failure for any injected \texttt{FF} faults in the design. While registers are simpler to ensure they are properly protected, netlist simulations inject faults in primitives, which include both combinational and sequential logic, resulting in a remaining failure rate of \SI{0.12}{\percent}. These faults can be traced back to components in the design that are not expected to be protected with the available methods, such as the \cgls{gpio} output signals, which are voted prior to connecting to the single output pad.

While injecting faults into the synthesized netlist provides a good representation of failure rates from the implemented logic, additional transistors, such as clock buffers, are not fully accounted for, but are sure to have a severe impact on the fault tolerance of the design.

\subsection{Who checks the checker?}
To verify the improvement in fault tolerance, we inject \texttt{PORT} faults into the outputs of the \cgls{tcls} majority voters and the outputs of the \cgls{ecc} decoders in the \cgls{rtl} simulation.

In Cfg.~2, the \cgls{tcls} majority voters do not overlap with the \cgls{ecc} decoders, located near the \cgls{sram} banks. Correspondingly, injecting faults results in \SI{16.33}{\percent} failures and \SI{1.59}{\percent} latent errors, with the remaining \SI{82.08}{\percent} of injected faults being masked.

In Cfg.~3, the \cgls{ecc} decoders are part of the \cgls{relOBI} protection, which overlaps with the \cgls{tcls} protection. Therefore, injecting errors in the \cgls{tcls} majority voters and the \cgls{relOBI} \cgls{ecc} decoders results in \SI{44.44}{\percent} of injected faults leading to a correction, and \SI{55.30}{\percent} of faults masking the injected error. The remaining \SI{0.26}{\percent} of simulations result in a failure, due to a mismatch in the \texttt{core\_busy\_o} voter output, a signal that is directly monitored as a chip \cgls{io} but is not otherwise used in the \cgls{soc}. This highlights that overlapping protection mechanisms even allow for the detection and correction of faults in the encoding, voting, and checking methods themselves.

\begin{figure}[t]
    \centering
    \includegraphics[width=0.2603\textwidth]{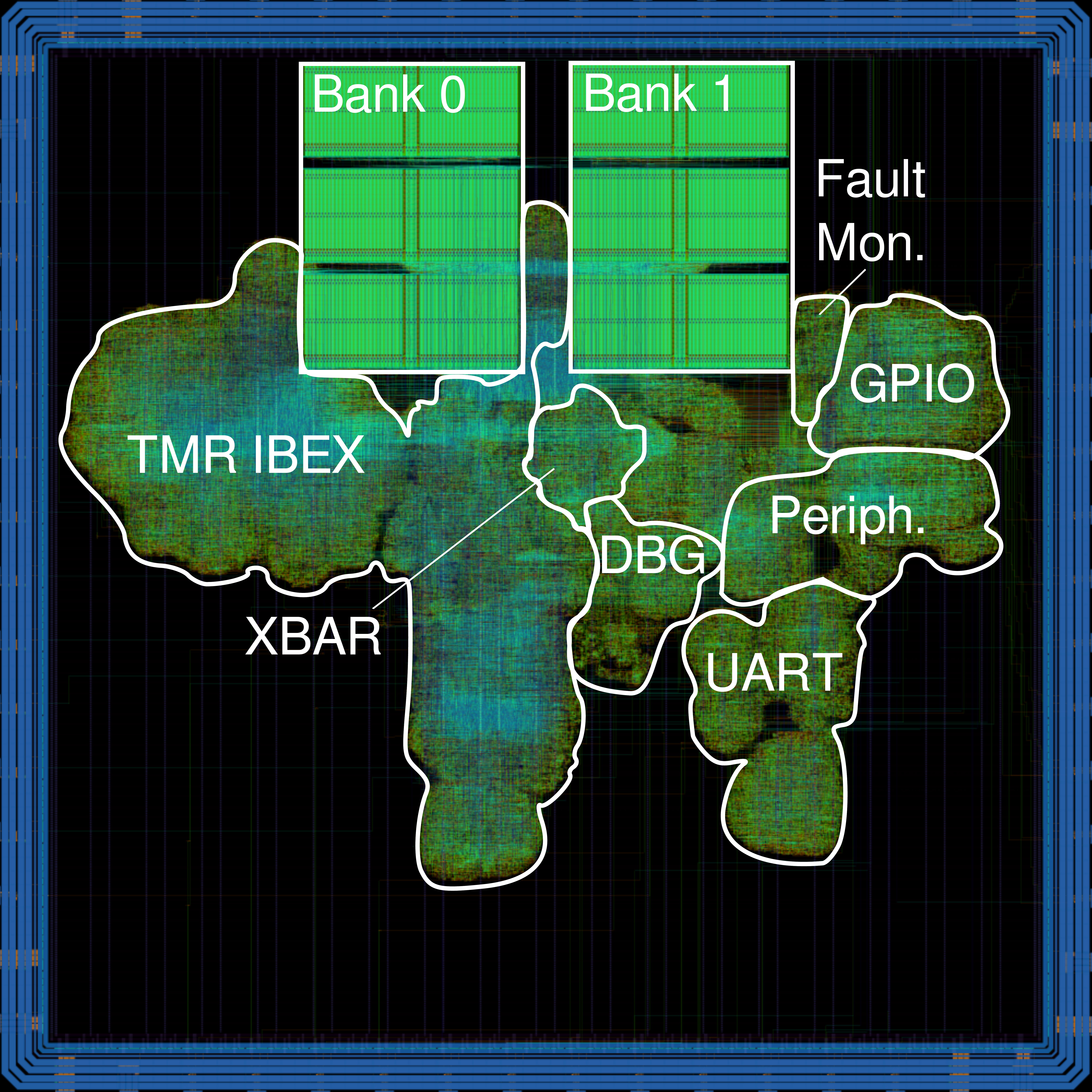}
    \caption{Annotated \textit{\cgls{tmrg}} layout render}
    \label{fig:tmrg_die}
    \vspace{-12pt}
\end{figure}

\section{Related Work}
\label{sec:related}

To compare \cgls{sota} solutions~\cite{andorno_radiation-tolerant_2023, walsemann_strv_2023} with our implementation on even ground, we implement \texttt{croc} with \cgls{tmrg}~\cite{kulis_single_2017}, triplicating all internal logic protected in Cfg.~4 and adding triplicated voters following every \cgls{ff}. The \cgls{sram} banks are triplicated as a whole, and the debug module is excluded from triplication, as in Cfg.~4.

The design protected with \cgls{tmrg} exhibits a \SI{1.28}{\times} higher area overhead than the fully protected design using individual protection mechanisms. While the total number of registers in the design is \SI{13}{\percent} lower, the additional combinational logic required for fine-grained triplication, and especially the fine interconnections required within, result in a more challenging design to implement and require a larger 4$\times$\SI{4}{\milli\meter\squared} die area, as shown in \cref{fig:tmrg_die}. Timing is also severely impacted due to the increased complexity of the design.

When injecting faults, shown in \cref{fig:measurement}, we see that all \cglspl{seu} result in a correct outcome, guaranteed by the voting directly after the registers. Latent errors are not measured, but due to the absence of a scrubber, faults in the \cgls{sram} banks are not expected to be corrected. Injecting \cglspl{set} in the synthesized netlist, we observe a \SI{0.12}{\percent} fault rate. Similar to Cfg.~4, these remaining faults are due to faults in the output voters of \cgls{io} pins, resulting in incorrect \cgls{io} behavior.

\subsection{Pareto Analysis of Design Configurations}

\begin{figure}[t]
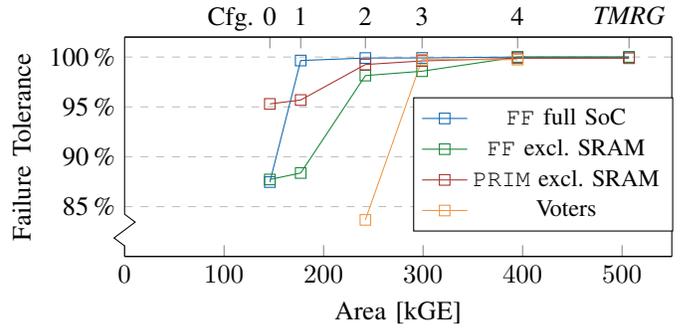

    \centering
    \include{fig/scatter.tex}
    \caption{Pareto-curves of fault-tolerance configurations}
    \label{fig:pareto}
    \vspace{-12pt}
\end{figure}

To contextualize our results, we define a Pareto frontier as the set of designs for which no lower area alternative configuration provides equal or higher fault tolerance across all fault scenarios. Applying this criterion to our design space, we show in \cref{fig:pareto} that all of the configurations adding tolerance (Cfgs.~1-4) lie along the Pareto-optimum.

 In contrast, fine-grained \cgls{tmr} across the complete design with \cgls{tmrg} lies strictly beyond the Pareto frontier: although it provides almost identical single-fault coverage to Cfg.~4, it requires $\sim$\SI{28}{\percent} additional area than Cfg.~4 and does not recover the frequency loss, thus offering no dominant improvement.

\section{Conclusion}

In summary, we show that individual protection methods can be combined to achieve similar fault-tolerance levels as \cgls{sota} solutions, such as fine-grained triplication of the entire \cgls{soc}, with lower area penalties. Properly overlapping the fault-tolerance encodings is crucial to ensure there are no gaps in protection on signals between the fault-tolerant domains or their voting, encoding, or decoding logic. Picking more appropriate methods can greatly improve implementability and reduce the area overhead by \SI{22}{\percent}.

Incrementally adding different protection methods furthermore allows for requirement balancing, trading off overhead with fault tolerance. Adding \cgls{ecc}-protected \cgls{sram} already addresses \SI{97}{\percent} of failures. Adding fault-tolerance with \cgls{tcls} for the processor cores addresses \SI{84}{\percent} of the remaining faults in the design. The final protection methods can achieve fault tolerance on par with \cgls{sota} at additional cost.

\clearpage

\bibliographystyle{IEEEtran}
\bibliography{style, references}

\end{document}

%% file: fig/scatter.tex
\begin{tikzpicture}
\pgfplotsset{set layers} 
\begin{axis}[
    xlabel={Area [\si{\kilo GE}]},
    ylabel={Failure Tolerance},
    xmin=0, xmax=550,
    ymin=80, ymax=102,
    ytick={85,90,95,100},
    yticklabels={\SI{85}{\percent},\SI{90}{\percent},\SI{95}{\percent},\SI{100}{\percent}},
    ymajorgrids=true,
    grid style=dashed,
    axis y discontinuity=crunch, 
    axis y line*=left,
    height=4.5cm,
    width=\columnwidth,
    legend style={font=\small},
    xtick pos = bottom,
    legend style={at={(axis cs:290,82.5)},anchor=south west},
    ytick style={draw=none},
]

\addplot[
    color=PULPBlue,
    mark=square,
    ]
    coordinates {
    (146,87.46)(177,99.65)(242,99.90)(299,99.91)(395,100)(507,100)
    };
\addplot[
    color=PULPGreen,
    mark=square,
    ]
    coordinates {
    (146,87.72)(177,88.37)(242,98.14)(299,98.58)(395,100)(507,100)
    };
\addplot[
    color=PULPRed,
    mark=square,
    ]
    coordinates {
    (146,95.3)(177,95.69)(242,99.25)(299,99.61)(395,99.88)(507,99.88)
    };
\addplot[
    color=PULPOrange,
    mark=square,
    ]
    coordinates {
    (242,83.67)(299,99.74)(395,99.74)
    };

    \legend{\texttt{FF} full \cgls{soc}, \texttt{FF} excl. \cgls{sram}, \texttt{PRIM} excl. \cgls{sram}, Voters}

\end{axis}
\begin{axis}[
      axis y line*=right,
      ymin=0, ymax=120,
      xmin=0, xmax=550,
      height=4.5cm,
      width=\columnwidth,
      ytick=\empty,
      xtick={146,177,242,299,395,507},
      xticklabels={{Cfg. 0\hphantom{Cfg. }},1,2,3,4,\textit{TMRG}},
      xtick pos=top,
      x tick label style={
        anchor=north,yshift=15pt}
    ]
    \end{axis}
\end{tikzpicture}
\vspace{-1.5cm}